\documentclass[letterpaper]{amsart}
\pdfoutput=1
\newtheorem{theorem}{Theorem}[section]

\newtheorem{corollary}[theorem]{Corollary}

\theoremstyle{definition}

\theoremstyle{remark}
\newtheorem{remark}[theorem]{Remark}
\usepackage{hyperref}
\usepackage{graphicx}
\usepackage{color}
\usepackage{amsmath}
\usepackage{amsfonts}
\usepackage{amssymb}
\usepackage{epsfig}
\usepackage{rotating}
\usepackage{graphics}
\usepackage{mathrsfs}
\newcommand{\be}{\begin{equation}}

\newcommand{\ee}{\end{equation}}

\newcommand{\ba}{\begin{array}}

\newcommand{\ea}{\end{array}}

\newcommand{\beq}{\begin{eqnarray}}

\newcommand{\eeq}{\end{eqnarray}}

\newtheorem{lm}{lemma}

\newtheorem{thee}{theorem}

\newtheorem{proo}{proposition}

\newtheorem{co}{corollary}

\newtheorem{rem}{remark}

\newtheorem{deff}{definition}

\newcommand{\bd}{\begin{deff}}

\newcommand{\ed}{\end{deff}}

\newcommand{\bl}{\begin{lm}}

\newcommand{\el}{\end{lm}}

\newcommand{\bp}{\begin{proo}}

\newcommand{\ep}{\end{proo}}

\newcommand{\bt}{\begin{thee}}

\newcommand{\et}{\end{thee}}

\newcommand{\bc}{\begin{co}}

\newcommand{\ec}{\end{co}}

\newcommand{\brm}{\begin{rem}}

\newcommand{\erm}{\end{rem}}

\newcommand{\der}{{\rm d}}

\usepackage{t1enc}
\def\frak{\mathfrak}

\newcommand{\newc}{\newcommand}

\let\ccdot\cdot
\def\cdot{\hbox to 2.5pt{\hss$\ccdot$\hss}}

\newc{\aR}{\mbox{\boldmath{$ R$}}}
\newc{\aS}{\mbox{\boldmath{$ S$}}}
\newc{\aT}{\mbox{\boldmath{$ T$}}}
\newc{\aW}{\mbox{\boldmath{$ W$}}}

\newc{\aK}{\mbox{\boldmath{$ K$}}}
\newc{\aL}{\mbox{\boldmath{$ L$}}}

\newcommand{\bbC}{\mathbb{C}}

\usepackage{amssymb}
\usepackage{amscd}

\newcommand{\bma}{\begin{pmatrix}}
\newcommand{\ema}{\end{pmatrix}}

\newc{\obstrn}[2]{B^{#1}_{#2}}

\newcommand{\rpl}                         
{\mbox{$
\begin{picture}(12.7,8)(-.5,-1)
\put(0,0.2){$+$}
\put(4.2,2.8){\oval(8,8)[r]}
\end{picture}$}}

\newcommand{\lpl}                         
{\mbox{$
\begin{picture}(12.7,8)(-.5,-1)
\put(2,0.2){$+$}
\put(6.2,2.8){\oval(8,8)[l]}
\end{picture}$}}

\usepackage{ifthen}

\newc{\tensor}[1]{#1}
\newc{\Mvariable}[1]{\mbox{#1}}
\newc{\down}[1]{{}_{#1}}
\newc{\up}[1]{{}^{#1}}

\newc{\JulyStrut}{\rule{0mm}{6mm}}
\newc{\midtenPan}{\mbox{\sf S}}
\newc{\midten}{\mbox{\sf T}}
\newc{\midtenEi}{\mbox{\sf U}}
\newc{\ATen}{\mbox{\sf E}}
\newc{\BTen}{\mbox{\sf F}}
\newc{\CTen}{\mbox{\sf G}}

\def\sideremark#1{\ifvmode\leavevmode\fi\vadjust{\vbox to0pt{\vss
 \hbox to 0pt{\hskip\hsize\hskip1em
 \vbox{\hsize3cm\tiny\raggedright\pretolerance10000
 \noindent #1\hfill}\hss}\vbox to8pt{\vfil}\vss}}}%

\numberwithin{equation}{section}

\newcounter{romenumi}
\newcommand{\labelromenumi}{(\roman{romenumi})}

\begin{document}
\title{Radiative Poincare type eon and its follower}
\vskip 1.truecm
\author{Pawe\l~ Nurowski} \address{Centrum Fizyki Teoretycznej, Polska Akademia Nauk, Al. Lotnik\'ow 32/46, 02-668 Warszawa, Poland}
\email{nurowski@cft.edu.pl}
\thanks{The research was funded from the Norwegian Financial Mechanism 2014-2021 with project registration number 2019/34/H/ST1/00636.}
\begin{abstract}
  We consider two consecutive eons $\hat{M}$ and $\check{M}$ from Penrose's Conformal Cyclic Cosmology and study how the matter content of the past eon ($\hat{M}$) determines the matter content of the present eon ($\check{M}$) by means of the reciprocity hypothesis.

  We assume that the only matter content in the final stages of the past eon is a spherical wave described by Einstein's equations with the pure radiation energy momentum tensor $$\hat{T}^{ij}=\hat{\Phi}K^iK^j,\quad\hat{g}_{ij}K^iK^j=0,$$ 
and with cosmological constant $\hat{\Lambda}$ . We solve these Einstein's equations associating to $\hat{M}$ the metric $\hat{g}=t^{-2}\big(-\der t^2+h(t)\big)$, which is a Lorentzian analog of the Poincar\'e-Einstein metric known from the theory of conformal invariants. The solution is obtained under the assumption that the 3-dimensional conformal structure $[h]$ on the $\mathscr{I}^+$ of $\hat{M}$ is flat, that the metric $\hat{g}$ admits a power series expansion in the time variable $t$, and that $h(0)\in [h]$. Such solution depends on one real arbitrary function of the radial variable $r$.

Applying the reciprocal hypothesis, $\hat{g}\to \check{g}=t^4\hat{g}$, we show that the new eon $(\check{M},\check{g})$ created from the one containing a single spherical wave, is filled at its initial state with three types of radiation: (i) the damped spherical wave which continues its life from the previous eon, (ii) the in-going spherical wave obtained as a result of a collision of the wave from the past eon with the Bang hypersurface and (3) randomly scattered waves that could be interpreted as perfect fluid with the energy density $\check{\rho}$ and the isotropic pressure $\check{p}$ such that $\check{p}=\tfrac13\check{\rho}$. The metric $\check{g}$ solves the Einstein's equations without cosmological constant and with the energy-momentum tensor
$$\check{T}^{ij}=\check{\Phi} K^iK^j+\check{\Psi}L^iL^j+(\check{\rho}+\check{p})\check{u}^i\check{u}^j+\check{p}\check{g}^{ij},$$
in which $\check{u}^i\check{u}^j\check{g}_{ij}=-1$, $\check{g}_{ij}L^iL^j=0$ and  $L^iK^j\check{g}_{ij}=-2$. 
  \end{abstract}

\date{\today}
\maketitle
\newcommand{\bbS}{\mathbb{S}}
\newcommand{\bbR}{\mathbb{R}}
\newcommand{\sog}{\mathbf{SO}}
\newcommand{\glg}{\mathbf{GL}}
\newcommand{\slg}{\mathbf{SL}}
\newcommand{\og}{\mathbf{O}}
\newcommand{\soa}{\frak{so}}
\newcommand{\gla}{\frak{gl}}
\newcommand{\sla}{\frak{sl}}
\newcommand{\sua}{\frak{su}}
\newcommand{\dr}{\mathrm{d}}
\newcommand{\sug}{\mathbf{SU}}
\newcommand{\gat}{\tilde{\gamma}}
\newcommand{\Gat}{\tilde{\Gamma}}
\newcommand{\thet}{\tilde{\theta}}
\newcommand{\Thet}{\tilde{T}}
\newcommand{\rt}{\tilde{r}}
\newcommand{\st}{\sqrt{3}}
\newcommand{\kat}{\tilde{\kappa}}
\newcommand{\kz}{{K^{{~}^{\hskip-3.1mm\circ}}}}
\newcommand{\bv}{{\bf v}}
\newcommand{\di}{{\rm div}}
\newcommand{\curl}{{\rm curl}}
\newcommand{\cs}{(M,{\rm T}^{1,0})}
\newcommand{\tn}{{\mathcal N}}
\newcommand{\ten}{{\Upsilon}}
\noindent
\section{The setting}
In this short note we show a model of a bandage region of two consecutive eons from the \emph{Penrose's Conformal Cyclic Cosmology} (CCC) \cite{pen}, which have the following properties\footnote{We closely follow Paul Tod's setting, notation and terminology as presented in \cite{pt}}:
\begin{itemize}
\item The common three-surface $\Sigma$ of $\mathscr{I}^+$ of the past eon and the Big Bang of the present eon is equipped with a conformal class $[h_0]$ of signature $(+,+,+)$ which \emph{has vanishing Cotton tensor}, i.e. $[h_0]$ is \emph{conformally flat}; in the following we will chose $h_0$ to be the \emph{flat} representative of the conformal class $[h_0]$;
\item The Poincar\'e-type extension $(\hat{M},\hat{g})$, with $\hat{M}=]0,\epsilon[\times \Sigma$, of the conformal three-manifold $(\Sigma,[h_0])$ has the Lorentzian metric \cite{fef}:
  \be \hat{g}=t^{-2}(-\der t^2+h_t);\label{gcheck}\ee
  Here $t\in]0,\epsilon[$ is a coordinate along the extension $\bbR_+$ of $\Sigma$ in $\hat{M}$, and $h_t=h(t,x)$ is a $t$-dependent 1-parameter family of metrics on $\Sigma$ such that $h(0,x)=h_0\in[h_0]$; here $x$ is a point in $\Sigma$, $x\in\Sigma$;
  \item The Poincar\'e-type metric $\hat{g}$ in $\hat{M}$ satisfies the \emph{pure radiation Einstein's equations with cosmological constant} $\hat{\Lambda}$:
    \be\hat{R}^{ij}=\hat{\Lambda} \hat{g}^{ij}+\hat{\Phi} K^i K^j;\label{eincheck}\ee
    Here $K^i$ is an \emph{expanding null} vector field \emph{without shear and without twist} on $\hat{M}$; in particular we have $\hat{g}_{ij}K^iK^j=0$;
  \item The Lorentzian four-metric $$g=-\der t^2+h_t$$ conformal to the Poincare-type metric $\hat{g}$ in $\hat{M}$ is naturally extended to $M=\hat{M}\cup\Sigma\cup\check{M}$, which is a bundle $\Sigma\to M\stackrel{\pi}{\to} I$ over the interval $I=]-\epsilon,\epsilon[\subset\bbR$ parameterized by $t$, with the following preimages of $I$: $\pi^{-1}(t>0)=\hat{M}$, $\pi^{-1}(t=0)=\Sigma$, and $\pi^{-1}(t<0)=\check{M}$;
  \item The metric $g$ is used to define a Lorentzian metric $\check{g}$ in $\check{M}$, which is
    $$\check{g}=t^2(-\der t^2+h_t),\quad\quad \mathrm{for}\quad t<0;$$
  \item Note that for $t>0$ we have $\hat{g}=\hat{\Omega}^2g$ and that for $t<0$ we have $\check{g}=\check{\Omega}^2g$ with $\check{\Omega}=-\hat{\Omega}^{-1}=t$.
  \end{itemize}

One of the aims of this note is to identify the above four-manifold $M$, equipped with the three Lorentzian metrics $\hat{g}$, $g$ and $\check{g}$, with the \emph{bandage region} \cite{pt} of the \emph{Penrose's cyclic Universe} \cite{pen} in which the \emph{past eon ends} as filled \emph{with only one spherical wave}\footnote{We do not specify what kind of wave it is: it may consists of the incoherent superposition of waves representing directed massless radiation with random phases and polarizations, but the same propagation direction $K^j$.} propagating along the null vector $K^i$. Forcing the Poincar\'e-type expansion metric $\hat{g}$ to satisfy the Einstein's equations \eqref{eincheck} is the first step to achieve this aim. Another aim is to see how the wave contained in the past eon will change into a matter content at the beginning of the present eon, by means of \emph{Penrose's reciprocal hypothesis}, stating that the three metrics $\hat{g}$, $g$ and $\check{g}$ in the bandage region should be related via: $\hat{g}=\Omega^{-2}g$ and $\check{g}=\Omega^2g$.

As we show below, under further simplification assumptions, the explicit form of the Poincar\'e-type metric $\hat{g}$ can be easily found up to an arbitrarily  prescribed accuracy, and as a byproduct one gets a \emph{remarkably pleasant} consequences of the so obtained $\hat{g}$ for $\check{g}$, and in particular for the \emph{matter content of the spacetime} $(\check{M},\check{g})$, which is interpreted as the beginning of the \emph{present eon}. 
\section{The ansatz and the model for the past eon}
In the \emph{theory of conformal invariants} as presented by Fefferman and Graham in \cite{fef}, given a conformal class $[h]$ on $\Sigma$, one obtains the system of conformal invariants of $[h]$ in terms of the (pseudo)Riemannian invariants of a certain (pseudo)Riemannian metric $\hat{g}$. This metric is naturally associated with the conformal class $[h]$ via  
$\hat{g}=t^{-2}(\varepsilon \der t^2+h_t)$, where $h_t$ is a 1-parameter family of metrics on $\Sigma$, such that $h_0=h$ and $h$ is a representative of $[h]$. The metric $\hat{g}$ is defined for $t>0$ and the value $\varepsilon=1$ is chosen. To encode the conformal properties of $[h]$, this metric is demanded to be \emph{unique}. This is done by the requirements that $\hat{g}$ is \emph{Einstein}, $\hat{Ric}(\hat{g})=\hat{\Lambda}\hat{g}$, and that $h_t$ is real analytic and symmetric in $t$.

We present here a \emph{milder version of this construction}, applied to the conformally flat Riemannian structure $(\Sigma,[h])$ on a three-dimensional $\Sigma$, to obtain the metric $\hat{g}$ of the past eon with a desirable physical properties. In our `milder version' we do as follows: \begin{itemize}
\item[(a)] We replace $\varepsilon=1$ by $\varepsilon=-1$ - to have the Lorentzian signature of $\hat{g}$; \item[(b)] We replace the \emph{Einstein condition} $\hat{Ric}(\hat{g})=\hat{\Lambda}\hat{g}$ by the Einstein equation \eqref{eincheck} - to have the past eon filled by a spherical wave; 
  \item[(c)] And we drop the condition that $h_t$ is symmetric in the variable $t$ - to have more flexibility on the matter content of the present eon $\check{M}$. 
\end{itemize}

Since in our milder-than-Fefferman-Graham-setting we do not have uniqueness theorems as in [1] the outcome past eon metric $\hat{g}$ is not rigidly constrained. Thus, instead of working with the most general form of the the 1-parameter family of metrics $h_t$ \emph{we make a physically motivated ansatz} for them, hoping that it is compatible with the Einstein's equations \eqref{eincheck}.

We start with a conformal class $[h_0]$ represented by the flat 3-dimensional metric
$$h_0=\frac{2r^2\der z \der\bar{z}}{(1+\frac{z\bar{z}}{2})^2}+\der r^2.$$
Then as $h_t$ we take the \emph{spherically symmetric} 1-parameter family
$$h_t=\frac{2r^2\big(1+\nu(t,r)\big)\der z \der\bar{z}}{(1+\frac{z\bar{z}}{2})^2}+\big(1+\mu(t,r)\big)\der r^2,$$
where the unknown function $\nu=\nu(t,r)$ and $\mu=\mu(t,r)$ are both \emph{real analytic} in the variable $t$ and such that:
$$\nu(0,r)=0\quad\mathrm{and}\quad\mu(0,r)=0.$$
This obviously satisfies $h_{t=0}=h_0$ and because of the analyticity assumption we have
$$\nu(t,r)=\sum_{i=1}^\infty a_i(r) t^i\quad\mathrm{and}\quad \mu(t,r)=\sum_{i=1}^\infty b_i(r) t^i,$$
with a set of differentiable functions $a_i=a_i(r)$ and $b_i=b_i(r)$ depending on the $r$ variable only.

This leads to the following ansatz for the pre-Poincar\'e-type metric $\hat{g}$ in $\hat{M}$:
\be\hat{g}=t^{-2}\,\Big(\,-\der t^2\,+\,\frac{2r^2\big(\,1+\sum_{i=1}^\infty a_i(r) t^i\,\big)\der z \der\bar{z}}{(1+\frac{z\bar{z}}{2})^2}\,+\,\big(\,1+\sum_{i=1}^\infty b_i(r) t^i\,\big)\der r^2\,\Big).\label{prem}\ee
Our (pre)past eon manifold $\hat{M}$ is parameterized by $t>0$, $r>0$ and $z\in\bbC\cup\{\infty\}$.

We now consider the following null vector field $K$ on $\hat{M}$:
$$K=\partial_t+\Big(\,1+\sum_{i=1}^\infty b_i(r) t^i\,\Big)^{-\tfrac12}\partial_r.$$
It is tangent to a congruence of null geodesics without shear and twist, which represents light rays emanating from the source at the surface $r=0$. We require that the Poincar\'e-type metric \eqref{prem} satisfies the Einstein equations \eqref{eincheck} with this null vector field $K$ and some functions $\hat{\Phi}$ and $\hat{\Lambda}$. We have the following theorem.\\

\noindent
{\bf Theorem 1.}\\
If the metric 
\be\begin{aligned}
  \hat{g}=&t^{-2}(-\der t^2+h_t)=\\&
  t^{-2}\,\Big(\,-\der t^2\,+\,\frac{2r^2\big(\,1+\nu(t,r)\,\big)\der z \der\bar{z}}{(1+\frac{z\bar{z}}{2})^2}\,+\,\big(\,1+\mu(t,r)\,\big)\der r^2\,\Big)=
  \\&t^{-2}\,\Big(\,-\der t^2\,+\,\frac{2r^2\big(\,1+\sum_{i=1}^\infty a_i(r) t^i\,\big)\der z \der\bar{z}}{(1+\frac{z\bar{z}}{2})^2}\,+\,\big(\,1+\sum_{i=1}^\infty b_i(r) t^i\,\big)\der r^2\,\Big)\end{aligned}\label{solcheck}\ee
satisfies Einstein's equations
\be \hat{E}{}_{ij}:=\hat{R}{}_{ij}-\hat{\Lambda}\hat{g}{}_{ij}-\hat{\Phi}\hat{K}{}_i\hat{K}{}_j=0\label{eiwei}\ee
with
\be K=K^i\partial_i=\partial_t+\Big(\,1+\sum_{i=1}^\infty b_i(r) t^i\,\Big)^{-\tfrac12}\partial_r,\quad\quad\hat{K}_i=\hat{g}_{ij}K^j, \label{eiwei1}\ee
then we have:
\begin{itemize}
\item The coefficients $a_1(r)$, $a_2(r)$ $b_1(r)$ and $b_2(r)$ identically vanish, $a_1(r)=a_2(r)=b_1(r)=b_2(r)=0$, and the power series expansion of $h_t$ starts at the $t^3$ terms, $h_t=t^3\chi(r)+\mathcal{O}(t^4)$.
\item The metric $\hat{g}$, or what is the same, the power series expansions $\nu(t,r)=\sum_{i=1}^\infty a_i(r) t^i$ and $\mu(t,r)=\sum_{i=1}^\infty b_i(r) t^i$, are totally determined up to infinite order by an arbitrary differentiable function $f=f(r)$.
\item More precisely, the Einstein equations $\hat{E}{}_{ij}=\mathcal{O}(t^{k+1})$ solved up to an order $k$, together with an arbitrary differentiable function $f=f(r)$,  uniquely determine $\nu(t,r)$ and $\mu(t,r)$ up to the order $(k+2)$.
  \item In the lowest order the solution reads:
    $$\nu=\frac{f}{r^3}t^3+\mathcal{O}(t^4)\quad\mathrm{ and}\quad \mu=-\frac{2f}{r^3}t^3+\mathcal{O}(t^4);$$
    The energy function $\hat{\Phi}$ and the cosmological constant $\hat{\Lambda}$ are:
    $$\hat{\Phi}=3\frac{f'}{r^3}t^6+\mathcal{O}(t^7)\quad\mathrm{ and}\quad \hat{\Lambda}=3+\mathcal{O}(t^{4});$$ the Weyl tensor of the solution is
    $$\hat{W}^i{}_{jkl}=\mathcal{O}(t).$$ In particular, the Weyl tensor $\hat{W}^i{}_{jkl}$ vanishes at $t=0$ and $\hat{\Lambda}>0$ there. 
\end{itemize}

\vspace{0.5cm}
\noindent
With the use of computers we calculated this solution up to the order $k=10$, finding explicitly $\nu=\sum_{k=3}^{10}a_kt^k$ and $\mu=\sum_{k=3}^{10}b_kt^k$. The formulas are compact enough up to $k=8$ and up to the order $k=8$ they read:
  $$\begin{aligned}
  \nu(t,r)=&f\tfrac{t^3}{r^3}\,-\,\tfrac34f'\tfrac{t^4}{r^4}+\tfrac{1}{10}\big(-2rf'+3r^2f''\big)\tfrac{t^5}{r^5}+\\&\tfrac{1}{24}\big(3f^2-3rf'+3r^2f''-2r^3f^{(3)}\big)\tfrac{t^6}{r^6}+\\
  &\tfrac{r}{280}\big(-24f'-105ff'+24rf''-12r^2f^{(3)}+5r^3f^{(4)}\big)\tfrac{t^7}{r^7}-\\&
  \tfrac{r}{960}\big(60f'+288ff'-150rf'{}^2-60rf''-216rff''+30r^2f^{(3)}-10r^3f^{(4)}+3r^4f^{(5)}\big)\tfrac{t^8}{r^8}
 +\\&\mathcal{O}(\big(\tfrac{t}{r}\big)^{9})
\end{aligned}
$$
\vspace{0.5cm}
  $$\begin{aligned}
  \mu(t,r)=&-2f\tfrac{t^3}{r^3}\,+\,\tfrac34f'\tfrac{t^4}{r^4}-\tfrac{1}{5}f''\tfrac{t^5}{r^5}\,+\,\tfrac{1}{24}\big(39f^2+r^3f^{(3)}\big)\tfrac{t^6}{r^6}\,-\,\tfrac{r}{280}\big(390ff'+2r^3f^{(4)}\big)\tfrac{t^7}{r^7}+\\&
  \tfrac{r}{960}\big(-18ff'+300rf'{}^2+378rff''+r^4f^{(5)}\big)\tfrac{t^8}{r^8}+\mathcal{O}(\big(\tfrac{t}{r}\big)^{9}).
\end{aligned}
  $$
For a solution up to this order we find that:
$$\scriptsize{\begin{aligned}
  \hat{\Phi}\,=\,&3r^3f'\tfrac{t^6}{r^6}\,+\,3r^3\big(f'-rf''\big)\tfrac{t^7}{r^7}\,+\,\tfrac{3r^3}{2}\big(2f'-2rf''+r^2f^{(3)}\big)\tfrac{t^8}{r^8}\,+\\&\tfrac{r^3}{2}\big(6f'+6ff'-6rf''+3r^2f^{(3)}-r^3f^{(4)}\big)\tfrac{t^9}{r^9}+
  \\&\tfrac{r^3}{8}\big(24f'+66ff'-12rf'{}^2-24rf''-30rff''+12r^2f^{(3)}-4r^3f^{(4)}+r^4f^{(5)}\big)\tfrac{t^{10}}{r^{10}}+
  \\&\tfrac{r^3}{40}\big(120f'+522ff'-177rf'{}^2-120rf''-378rff''+93r^2f'f''+60r^2f^{(3)}+90r^2ff^{(3)}-20r^3f^{(4)}+5r^4f^{(5)}-r^5f^{(6)}\big)\tfrac{t^{11}}{r^{11}}+\\&\mathcal{O}(\Big(\tfrac{t}{r}\Big)^{12}),
  \end{aligned}}$$
$$\hat{\Lambda}=3+\mathcal{O}(t^9).$$
I have no patience to type the Weyl tensor components up to high order. It is enough to say that that up to the 4th order in $t$,  modulo a nonzero constant tensor $C^i{}_{jkl}$, it is equal to:
$$\hat{W}^i{}_{jkl}=\Big(\frac{f}{r^2}\frac{t}{r}-\frac{f'}{r}\frac{t^2}{r^2}+\frac{f''}{2}\frac{t^3}{r^3}\Big) C^i{}_{jkl}+\mathcal{O}(\Big(\tfrac{t}{r}\Big)^4).$$

Of course, for the positivity of the energy density $\hat{\Phi}$ close to the surface $\mathscr{I}^+$ of $\hat{M}$ we need
$$f'>0.$$
\begin{remark}
  We were unable to find a reccurence relation for the functions $a_i(r)$ and $b_i(r)$ for arbitrary $i>10$. We nevertheless claim that such relations do exist  and that the corresponding power series are convergent. The reson for these claims is that our solution for $\hat{g}$ is a pure radiation Einstein metric with cosmological constant, which have a sherafree expanding but notwisting congruence of null geodesics which is tangent to the wave propagation vector $K^i$. All such solutions of Einstein's equations are known. They belong to the Robinson-Trautman class of solutions described e.g. in Chapter 28.4 of Ref. \cite{kramer}. Our solution is the spherically symmetric solution from this class \cite{vaidya}, and can be written in terms of the Robinson-Trautman coordinates \cite{RT} as:
  \be \hat{g}=\frac{2v^2\der\zeta\der\bar{\zeta}}{(1+\tfrac12\zeta\bar{\zeta})^2}-2\der u\big(\der v+(1-\frac{2m(u)}{v}-\tfrac13\Lambda v^2)\der u\big).\label{cci}\ee
  The trouble is that, because of the appearence of the free funcion $m=m(u)$ in \eqref{cci}, there is no an easy way of getting the explicit coordinate transformation from the Robinson-Trautman \emph{null} coordinates $(\zeta,\bar{\zeta},v,u) $ to our coordinates $(z,\bar{z},r,t)$. Such transformation would bring $\hat{g}$ as in \eqref{cci} to ours $\hat{g}$ from \eqref{prem} in which all the coefficients $a_i$ and $b_i$ are determined up to infinite order. Anyhow, knowing this transformation or not, the geometric features of our solution with all $a_i$s and $b_i$s determined for $i\to\infty$, show that our $\hat{g}$  must be identified with the Vaidya solution \eqref{cci}. Thus not only the coefficients $a_i$ and $b_i$ in our solution are determined up to infinite order, but also the power series defining our $\hat{g}$ \emph{converges} to $\hat{g}$ given by \eqref{cci}.
   \end{remark}
\begin{corollary}
The Poincar\'e-type metric \eqref{solcheck} can be interpreted as the ending stage of the evolution of the past eon in Penrose's CCC. The eon has a positive cosmological constant $\hat{\Lambda}\simeq 3$, which is filled with a spherically symmetric pure radiation moving along the null congruence generated by the vector field $K$.  
\end{corollary}

\section{Using reciprocity for the model of the present eon}
Now, following the Penrose-Tod reciprocal hypothesis procedure, we summarize the properties of the spacetime $\check{M}$ equipped with the metric $\check{g}$ obtained from $\hat{g}$ as in Theorem 1, by the reciprocal change $\check{\Omega}\to-\hat{\Omega}^{-1}=t$. 
In other words, we are now interested in the properties of the metric $\check{g}=t^4\hat{g}$. We have the following theorem.
\vspace{0.5cm}

\noindent
{\bf Theorem 2.}\\
Assume that the metric $\hat{g}$ as in \eqref{solcheck} satisfies the Einstein equations \eqref{eiwei}-\eqref{eiwei1}, $\hat{E}{}_{ij}=0$. Then, the reciprocal metric
$$\begin{aligned}
  \check{g}=&
  t^2\,\Big(\,-\der t^2\,+\,\frac{2r^2\big(\,1+\nu(t,r)\,\big)\der z \der\bar{z}}{(1+\frac{z\bar{z}}{2})^2}\,+\,\big(\,1+\mu(t,r)\,\big)\der r^2\,\Big)=
  \\&t^2\,\Big(\,-\der t^2\,+\,\frac{2r^2\big(\,1+\sum_{i=1}^\infty a_i(r) t^i\,\big)\der z \der\bar{z}}{(1+\frac{z\bar{z}}{2})^2}\,+\,\big(\,1+\sum_{i=1}^\infty b_i(r) t^i\,\big)\der r^2\,\Big)\end{aligned}$$
satisfies the Einstein equations
\be\check{E}{}_{ij}=\check{R}_{ij}-\check{\Phi}\check{K}_i\check{K}_j-\check{\Psi}\check{L}_i\check{L}_j-(\check{\rho}+\check{p})\check{u}_i\check{u}_j-\tfrac12(\check{\rho}-\check{p})\check{g}_{ij}=0.\label{eiwei2}\ee
Here $\check{K}_i$ and $\check{L}_i$ are the null 1-forms corresponding to the pair of {\color{green}out}going-{\color{red}in}going null vector fields $$K=K^i\partial_i=\partial_t{\color{green}+}\Big(\,1+\sum_{i=1}^\infty b_i(r) t^i\,\Big)^{-\tfrac12}\partial_r\quad\mathrm{ and}\quad L=L^i\partial_i=\partial_t{\color{red}-}\Big(\,1+\sum_{i=1}^\infty b_i(r) t^i\,\Big)^{-\tfrac12}\partial_r,$$ via $\check{K}_i=\check{g}_{ij}K^j$ and $\check{L}=\check{g}_{ij}L^j$, and the 1-form vector field $\check{u}_i$ corresponds to the future oriented\footnote{Note that now $t<0$ (!)} timelike unit vector field
$$\check{u}=\check{u}^i\partial_i=-t^{-1}\partial_t,$$
via $\check{u}_i=\check{g}_{ij}\check{u}^j$.

\vspace{0.7cm}
\noindent
Before giving the explicit formulas for the power expansions of functions $\check{\Phi}$, $\check{\Psi}$, $\check{\rho}$ and $\check{p}$ appearing in this theorem, we make the following remark.\\

\begin{remark}
The Einstein equations \eqref{eiwei2} are equations with an energy momentum tensor consisting of radiation propagating with spherical fronts outward (along $K$) and inward (along $L$); it also consists of a perfect fluid comoving with the present eon's cosmological time $T=-\int t\der t$. Each front of the spherical wave present in the past eon that reached the $t=0$ surface in the present eon produces (i) a \emph{spherical outward wave} with energy density $\check{\Phi}$ going along $K$ out of this sphere, (ii) a \emph{spherical inward wave} with energy density $\check{\Psi}$ going along $L$ \emph{towards the center of this sphere}, and (iii) a portion of a \emph{perfect fluid} with energy density $\check{\rho}$ and isotropic pressure $\check{p}$.  
\end{remark}
For the solutions $\nu(t,r)$, $\mu(t,r)$ of the past eon's Einstein's equations \eqref{eiwei}\eqref{eiwei1}, which were given in terms of the power series expansions as $\nu(t,r)=\sum_{i=3}^{k+2} a_i(r) t^i+\mathcal{O}(t^{k+3})$ and $\mu(t,r)=\sum_{i=3}^{k+2} b_i(r) t^i+\mathcal{O}(t^{k+3})$ in Theorem 1, the formulae for the power series expansions of the energy densities $\check{\Phi}$  $\check{\Psi}$, $\check{\rho}$ and the pressure $\check{p}$ are as follows:
$$\begin{aligned}
  \check{\Phi}=&-\frac{9f}{r^3}t^{-3}\,+\,\frac{9f'}{r^3}t^{-2}\,+\,\frac{1}{2r^4}\big(8f'-11rf''\big)t\,+\,\frac{3}{4r^5}\big(5f'-5rf''+3r^2f^{(3)}\big)\,+\\&\frac{9}{40r^6}\big(16f'+5ff'-16rf''+8r^2f^{(3)}-3r^3f^{(4)}\big)t\,+\\&
  \frac{1}{120r^7}\big(420f'+1068ff'-30rf'{}^2-420rf''-384rff''+210r^2f^{(3)}-70r^3f^{(4)}+19r^4f^{(5)}\big)t^2\,+\\&\dots+\mathcal{O}\big(t^{k-3}\big),
\end{aligned}$$
\vspace{0.5cm}
$$\begin{aligned}
  \check{\Psi}=&-\frac{9f}{r^3}t^{-3}\,+\,\frac{6f'}{r^3}t^{-2}\,+\,\frac{1}{2r^4}\big(2f'-5rf''\big)t^{-1}\,+\,\frac{3}{4r^5}\big(f'-rf''+r^2f^{(3)}\big)\,+\\&\frac{1}{40r^6}\big(24f'-75ff'-24rf''+12r^2f^{(3)}-7r^3f^{(4)}\big)t\,+\\&
  \frac{1}{60r^7}\big(30f'+39ff'+75rf'{}^2-30rf''+33rff''+15r^2f^{(3)}-5r^3f^{(4)}+2r^4f^{(5)}\big)t^2\,+\\&\dots+\mathcal{O}\big(t^{k-3}\big),
\end{aligned}$$
\vspace{0.5cm}
$$\begin{aligned}
  \check{\rho}=&3t^{-4}+\frac{18f}{r^3}t^{-1}\,-\,\frac{18f'}{r^3}\,+\,\frac{-6f'+9rf''}{r^4}t-\,\frac{3}{4r^6}\big(9f^2+3rf'-3r^2f''+2r^3f^{(3)}\big)t^2\,+\\&\frac{3}{20r^6}\big(-24f'+105ff'+24rf''-12r^2f^{(3)}+5r^3f^{(4)}\big)t^3\,-\\&
  \frac{1}{20r^7}\big(60f'+96ff'+120rf'{}^2-60rf''+72rff''+30r^2f^{(3)}-10r^3f^{(4)}+3r^4f^{(5)}\big)t^4\,+\\&\dots+\mathcal{O}\big(t^{k-1}\big),
\end{aligned}$$
\vspace{0.5cm}
$$\begin{aligned}
  \check{p}=&t^{-4}+\frac{6f}{r^3}t^{-1}\,+\,\frac{1}{r^4}\big(2f'-rf''\big)t+\,\frac{1}{2r^6}\big(18f^2+3rf'-3r^2f''+r^3f^{(3)}\big)t^2\,-\\&\frac{3}{20r^6}\big(-8f'+45ff'+8rf''-4r^2f^{(3)}+r^3f^{(4)}\big)t^3\,+\\&
  \frac{1}{30r^7}\big(30f'+57ff'+45rf'{}^2-30rf''+39rff''+15r^2f^{(3)}-5r^3f^{(4)}+r^4f^{(5)}\big)t^4\,+\\&\dots+\mathcal{O}\big(t^{k-1}\big).
  \end{aligned}$$
In these formulas all the \emph{doted} terms are explicitly determined in terms of $f$ and its derivatives (I was lazy, and typed only the terms adapted to the choice $k=6$ in Theorem 1).

The following remarks are in order:\\

\noindent
\begin{remark}~
\newline
\begin{itemize}
\item Note that since in $\check{M}$ the time  $t<0$, then the requirement that the energy densities are positive near the Big Bang hypersurface $t=0$ implies that
  $$f>0$$
  in addition to $f'>0$, which was the requirement we got from the past eon. Indeed, the leading terms in $\check{\Phi}$ and $\check{\Psi}$ are $\check{\Phi}=\check{\Psi}=-\frac{9f}{r^3}t^{-3}$, hence $\check{\Phi}$ and $\check{\Psi}$ are both positive in the regime $t\to 0^-$ provided that $f>0$. Note also that $f>0$ and $f'>0$ are the only conditions needed for the positivity of energy densities, as the leading term in $\check{\rho}$ is $\check{\rho}\simeq 3t^{-4}$, and is positive regardless of the sign of $t$.
\item Remarkably the leading terms in $\check{\rho}$ and $\check{p}$, i.e. the terms with negative powers in $t$, are proportional to each other with the numerical factor \emph{three}. We have
  $$\check{p}=\tfrac13\check{\rho}+\mathcal{O}(t^0).$$
  This means that immediately after the Bang, apart from the matter content of two spherical ingoing and outgoing waves in the new eon, there is also a scattered \emph{radiation} there, described by the perfect fluid with $\check{p}=\tfrac13\check{\rho}$.  \item So what the \emph{Penrose-Tod scenario does to the new eon out of a single spherical wave in the past eon}, is it splits this wave into \emph{three portions of radiation: the two spherical waves}, one which is a dumped continuation from the previous eon, the other that is focusing in the new eon, as it encountered a mirror at the Bang surface, \emph{and in addition a lump of scattered radiation described by the statistical physics}.
  
  \end{itemize}
\end{remark}

\end{document}